\documentclass[useamsfonts]{pasj02}
\usepackage{amsmath,amssymb,booktabs}
\usepackage{natbib}
\usepackage{url}
\newcommand{\href}[2]{#2}
\makeatletter
\def\ps@firstpage{%
  \def\@oddhead{\hbox to \textwidth{\footnotesize\itshape
    Accepted for publication in Publications of the Astronomical Society of Japan (2026)\hfil}}%
  \let\@evenhead\@oddhead
  \def\@oddfoot{\hfil\thepage\hfil}%
  \let\@evenfoot\@oddfoot
}
\makeatother
\jyear{2026}
\Received{}
\Accepted{}
\SetRunningHead{S. Abbassi and F. Habibi}{Angle-dependent emission from evolving accretion layers}
\begin{document}
\title{Angle-dependent emission from evolving accretion layers}
\author{Shahram \textsc{Abbassi}\altaffilmark{1}\altemailmark\email{sabbassi@uwo.ca}}
\author{Fahimeh \textsc{Habibi}\altaffilmark{2}\altemailmark\email{f.habibi@birjand.ac.ir}}
\altaffiltext{1}{Department of Physics and Astronomy, Western University, London, Ontario N6A 3K7, Canada}
\altaffiltext{2}{Department of Physics, University of Birjand, Birjand, Iran}
\maketitle
\begin{abstract}
Photon propagation through an evolving atmosphere can change the angular distribution of emerging radiation even when its integrated optical column remains constant. We quantify this effect in an irradiated, internally heated, expanding plane-parallel layer using grey, special-relativistic similarity solutions. The full angular radiation field determines both the source function and the acceleration of a pressureless flow in prompt radiative balance. We compare each evolving solution with two frozen transport calculations on the same density and velocity profiles, holding the source fixed in one and allowing it to relax in the other. Heating produces the familiar transition from limb darkening to limb brightening, while the temporal correction to the angular contrast changes sign with optical column. For an upper boundary moving at $0.04c$, an injection speed of $0.06c$, and heating four times the reference incident flux, the intensity ratio between direction cosines $0.1$ and $1$ exceeds the frozen, fixed-source result by $18.8$ per cent at column $\tau_0=0.15$, but is smaller by $2.53$ per cent at $\tau_0=0.55$. Source relaxation preserves this reversal. The corresponding direct photon flight times are one third and three per cent of the flow age in the oblique and normal directions, respectively. These solutions provide a controlled local benchmark for temporal angular transfer in evolving disc atmospheres and reprocessing layers; application to a specific accreting system requires thermal, spectral, and geometrical physics beyond the present model.
\end{abstract}

\section{Introduction}
Accretion onto black holes converts gravitational energy into radiation, but the emission received by a distant observer depends on how that radiation leaves the gas. The angular dependence of the emergent intensity, the geometry of the emitting region and relativistic propagation jointly determine the observed signal. In geometrically thin discs, atmosphere calculations and relativistic spectral models show why the emergent radiation must be related to the vertical structure and viewing inclination, rather than specified solely by the local energy flux \citep{Li2005,Davis2005,Davis2006}. Understanding the angular distribution of the escaping light is therefore an essential step in connecting accretion physics to observations.

This connection becomes more complex when the accretion flow develops a thick atmosphere or an outflow. At super-Eddington accretion rates, radiation transport and wind geometry can produce strongly inclination-dependent emission \citep{Poutanen2007,JiangGlobal2014}. Tidal disruption events provide a related setting in which the radiation is processed by gas whose density and geometry evolve as the accretion conditions change \citep{Dai2018,Thomsen2022}. Multidimensional transfer calculations further show that the available escape paths and ionization structure affect the emerging spectrum \citep{Parkinson2025}. These systems motivate a local question beneath the global geometrical problem: how does an evolving emitting layer redistribute radiation between different directions?

Even in a static atmosphere, the angular distribution depends on optical depth, illumination and the location of energy release. The classical semi-infinite atmosphere provides a familiar limb-darkened reference \citep{Chandrasekhar1960}. A finite layer can behave differently because an oblique ray samples a longer emitting path while attenuation and radiation incident from below compete with the distributed emission. Calculations for finite accretion-disc atmospheres consequently admit limb brightening as well as limb darkening \citep{FukueAkizuki2006,Fukue2011,Samadi2020}. Heating near disc surfaces has a physical basis: radiation magnetohydrodynamic simulations find that dissipation per unit mass can rise towards the upper layers \citep{JiangCorona2014}. The angle-dependent emission of heated, low-column gas therefore need not follow the law appropriate to an optically thick atmosphere in radiative equilibrium.

Bulk motion introduces a further dependence through aberration and the relativistic transformation of intensity, emission and extinction \citep{Mihalas1984,Fukue2014}. In a coupled flow, the resulting radiation field also supplies the force that changes the gas velocity, so transport and dynamics cannot generally be specified independently \citep{Fukue2015}. Temporal evolution adds another effect. When density and emissivity vary during photon passage, the escaping intensity depends on the conditions encountered along the ray, not only on the instantaneous atmospheric profiles. The relevant passage time is directional: oblique rays can sample appreciable evolution even when the vertical light-crossing time is short. A stationary angular law may then miss a contribution to the inclination dependence of the local emission.

Established methods provide several routes to this problem. Variable-Eddington-factor calculations determine the angular closure from the radiation field rather than imposing a universal relation between its moments \citep{Auer1970}. Relativistic transfer has been coupled to steady plane-parallel flows \citep{Fukue2015}, while similarity methods retain temporal transfer in prescribed relativistic flows \citep{Lucy2005} and in time-dependent plane-parallel radiation hydrodynamics \citep{Fukue2018}. Direct angular transport schemes offer a complementary approach, evolving intensities and evaluating radiation--matter exchange from their angular integrals \citep{Jiang2014,Jiang2021}. Alongside such general calculations, a controlled family of local solutions can help identify which changes in the emergent radiation arise specifically from temporal transport.

The question addressed here is whether evolution enhances oblique emission relative to the normal direction, or whether the sign of that correction depends on the optical column. Comparing unrelated steady and evolving atmospheres would mix this effect with differences in density, velocity and heating. We instead compare an evolving layer with frozen transport calculations on the same material profiles, retaining the velocity-dependent transfer coefficients. One control holds the source fixed, while a second allows the frozen radiation field and source to relax. This comparison tests whether a temporal change in angular contrast persists when the response of the source is included.

We construct a family of illuminated, internally heated, plane-parallel similarity solutions in which the full angular intensity determines both the source and the radiative acceleration. The similarity reduction retains photon flight times and radiation--matter exchange while allowing a survey of optical column, expansion rate and illumination. Within the adopted grey, cold-flow approximation, the gas velocity is a coupled unknown and no prescribed Eddington factor closes the transfer problem. Explicit conservation identities and independent numerical tests provide checks on the resulting solutions.

Section~2 specifies the radiating layer and its governing equations. Section~3 describes the coupled solution and the frozen comparisons, and Section~4 presents the angle-dependent emission and its numerical accuracy. Section~5 discusses the astrophysical applications and the domain of the model, followed by the conclusions in Section~6. Appendix~\ref{app:conservation} derives the conservation laws, and Appendix~\ref{app:analytic} gives the exact ray solutions used for verification.

\section{The radiating layer}
\subsection{An expanding surface layer}
We represent a small patch of an accretion atmosphere or reprocessing layer by a plane-parallel flow in flat spacetime, with laboratory time $t>0$, vertical coordinate $z$ and direction cosine $\mu$. The flow velocity is $\beta=\beta(\eta)=V/c$, with $\gamma(\eta)=[1-\beta(\eta)^2]^{-1/2}$. The grey extinction per unit proper rest mass, $\chi=\kappa+\sigma$, is constant. Subscript 0 denotes comoving radiation quantities. The similarity scalings are
\begin{align}
\eta&=z/(ct),&\rho_0&=R(\eta)/(\chi ct),\\
I(t,z,\mu)&=\frac{c^2}{4\pi\chi t}i(\eta,\mu),&
q^+&=\frac{c}{\chi t^2}q .\label{eq:scales}
\end{align}
Here $q^+$ is externally supplied energy per unit proper volume and proper time; $q$ is spatially uniform and constant in time within each model. Consequently, $q^+\propto t^{-2}$: uniform dimensionless heating does not mean constant dimensional heating. At fixed $\eta$, density and intensity decrease as $t^{-1}$ while the layer thickness grows as $t$. Section~\ref{sec:similarity} discusses the physical scope of this imposed evolution. Source intensities $S_0$ and $B_0$ use the same prefactor as $I$, defining $s_0$ and $b_0$. Material thermal inertia and pressure are neglected. Heat is immediately returned to the radiation as isotropic comoving emission, while the radiation force changes the bulk motion. This is a cold-flow, prompt-radiative-balance approximation, not a solution of a gas-temperature equation.

Define dimensionless laboratory moments
\begin{equation}
(j,h,k)=\frac12\int_{-1}^{1}(1,\mu,\mu^2)i\,d\mu .
\end{equation}
The corresponding energy density, flux and pressure are $c j/(\chi t)$, $c^2 h/(\chi t)$ and $c k/(\chi t)$. With
\begin{equation}
D=[\gamma(1-\beta\mu)]^{-1},\qquad
\mu_0=\frac{\mu-\beta}{1-\beta\mu},
\end{equation}
the comoving moments are
\begin{equation}
(j_0,h_0,k_0)=\frac12\int_{-1}^{1}(1,\mu_0,\mu_0^2)D^{-2}i\,d\mu .\label{eq:moments}
\end{equation}
The weight $D^{-2}$ includes both the bolometric intensity transformation and angular aberration. We calculate $f_0=k_0/j_0$ from these integrals; no angular closure is imposed.

\subsection{Heating and transfer}
For isotropic comoving redistribution, the frequency-integrated transfer equation is \citep{Mihalas1984}
\begin{equation}
\frac1c\frac{\partial I}{\partial t}+\mu\frac{\partial I}{\partial z}
=\rho_0\chi(D^3 S_0-D^{-1}I).
\end{equation}
With absorption fraction $\epsilon=\kappa/\chi>0$, the usual grey source is $s_0=(1-\epsilon)j_0+\epsilon b_0$. Prompt balance requires $\epsilon R(b_0-j_0)=q$, and hence
\begin{equation}
s_0=j_0+q/R.\label{eq:source}
\end{equation}
The elimination of $\epsilon$ is a consequence of the imposed energy balance, not a general equivalence of absorption and scattering. For any finite $\epsilon>0$, $b_0=j_0+q/(\epsilon R)$ gives the same bolometric source in equation~(\ref{eq:source}) at fixed $q$. Pure absorption means $\epsilon=1$, whereas $b_0=j_0$ denotes radiative equilibrium and implies $q=0$ in this model. With $q=0$, the balance source reduces to $s_0=j_0$, as it also does for conservative isotropic scattering. Strictly pure scattering ($\epsilon=0$) cannot return a nonzero imposed heat input to radiation through the absorption--emission term used here; a separate emissivity would be required. The limit $\epsilon\rightarrow0$ at fixed $q>0$ is therefore singular in $b_0$. ``Cold refers to neglected material pressure and thermal inertia, rather than zero temperature. For $q>0$, prompt balance includes net radiative cooling of the supplied heat and is not radiative equilibrium. The reduced grey source does not establish independence of realistic spectra from absorption. The adopted isotropic scattering law also excludes the angular and polarization structure of Thomson scattering.

Substitution of equation~(\ref{eq:scales}) gives
\begin{equation}
(\mu-\eta)\frac{\partial i}{\partial\eta}
=i+R(D^3s_0-D^{-1}i),\label{eq:transfer}
\end{equation}
or, equivalently,
\begin{equation}
\frac{\partial}{\partial\eta}[(\mu-\eta)i]
=R(D^3s_0-D^{-1}i).\label{eq:conservative}
\end{equation}
The conservative form avoids dividing the numerical equations by a vanishing characteristic speed at $\mu=\eta$. A finite differentiable solution there must satisfy $(R/D-1)i=RD^3s_0$. Conservative discretization handles the vanishing transport coefficient, but does not establish regularity for arbitrary parameter choices. All angular sectors enter equation~(\ref{eq:moments}). Discarding near-horizontal directions would change the source and force even if only more nearly vertical emergent rays were reported.

\subsection{Radiation and acceleration}
The gas exchanges momentum with the radiation while conserving rest mass. The equations below follow from the covariant conservation laws \citep{Mihalas1984}; Appendix~\ref{app:conservation} derives each term in the present scaling. Continuity gives the constant similarity mass flux
\begin{equation}
C_m=\gamma R(\beta-\eta).\label{eq:mass}
\end{equation}
For the prescribed positive mass flux $C_m>0$, positive density requires $\beta>\eta$. This selects gas moving upward relative to surfaces of fixed $\eta$; it is not a restriction on all possible similarity flows. The cold material acceleration under the comoving radiation flux obeys
\begin{equation}
\gamma^3(\beta-\eta)\frac{d\beta}{d\eta}=h_0.\label{eq:flow}
\end{equation}
Positive comoving flux accelerates the gas upwards; downward flux decelerates the injected flow. Evaluating this force from equation~(\ref{eq:moments}) keeps the dynamics tied to the radiation that forms the emergent intensity.

The angular collision integrals yield
\begin{align}
(h-\eta j)'&=\gamma(q-R\beta h_0),\label{eq:radenergy}\\
(k-\eta h)'&=\gamma(\beta q-Rh_0),\label{eq:radmomentum}
\end{align}
where a prime denotes $d/d\eta$. Combining these relations with equations~(\ref{eq:mass}) and (\ref{eq:flow}) gives the two useful checks
\begin{align}
[h-\eta j+C_m(\gamma-1)]'&=\gamma q,\label{eq:energy}\\
[k-\eta h+C_m\gamma\beta]'&=\gamma\beta q.\label{eq:momentum}
\end{align}
Subtracting the constant rest-mass contribution in equation~(\ref{eq:energy}) leaves the kinetic-energy flux. The right-hand sides include the laboratory energy and momentum deposited by isotropic comoving heating. They express the partition of supplied heat between escaping radiation and material kinetic energy, and provide direct checks of that partition in the numerical solution.

\subsection{Illumination and column}
The emitting patch occupies $\eta_b\leq\eta\leq\eta_s$. At the lower boundary we prescribe $\beta_b$ and incoming laboratory intensity
\begin{equation}
i_b(\mu)=i_*\frac{1+a\mu}{1+2a/3},\qquad \mu>\eta_b.\label{eq:bc}
\end{equation}
At the upper boundary, incoming intensity is zero for $\mu<\eta_s$.

These angular intervals follow from photon propagation relative to the moving boundaries. A photon has vertical laboratory speed $c\mu$, whereas the boundary at fixed $\eta$ moves at $c\eta$. Thus $c(\mu-\eta_b)>0$ identifies rays entering through the lower boundary, and $c(\mu-\eta_s)<0$ identifies rays entering through the upper boundary. The lower condition represents illumination from deeper emitting layers; the upper condition represents an exterior with negligible incident radiation. The latter includes shallow upward laboratory rays that the upper boundary overtakes. Outgoing intensities at both boundaries are determined by the solution. Zero upper inflow therefore does not mean zero emergent flux. The angular sectors are required by the characteristics, but the prescribed incident intensities are physical boundary assumptions.

 The reference illumination flux $h_*=i_*/4$ is the positive-hemisphere laboratory flux of equation~(\ref{eq:bc}); it is not the net moving-boundary flux. We define
\begin{equation}
\tau_0=\int_{\eta_b}^{\eta_s}R\,d\eta,\qquad
\mathcal H=\frac{q(\eta_s-\eta_b)}{h_*} .
\end{equation}
$\tau_0$ is a proper-opacity column, not the direction-dependent optical depth along a moving ray.

The reference parameters are $\eta_b=0.01$, $\eta_s=0.04$, $\beta_b=0.06$, $C_m=0.6$, $i_*=0.03$ and $a=0$. We study $\mathcal H=0,1,4$; a second series fixes $\mathcal H=4$ and adjusts $C_m$ to match selected columns. The boundaries delimit the part of the expanding atmosphere represented by the model. In particular, $\beta_s$ need not equal $\eta_s$: the upper boundary is not a material surface or a self-consistently located photosphere. Its exterior is treated as optically negligible.

The choice $\eta_b=0.01$ locates the base within an expanding, already moving layer above the unresolved launching region; its particular nonzero value is not essential to the formulation. A stationary lower boundary, $\eta_b=0$, is compatible with this branch provided $\beta_b>0$. In contrast, setting both $\eta_b=0$ and $\beta_b=0$ gives $C_m=0$ for finite base density and cannot supply the prescribed $C_m>0$. Approaching $\beta_b=\eta_b$ at fixed positive $C_m$ instead makes the density singular. A regular launch from rest would therefore require different boundary data or additional launching physics.

An unilluminated, internally heated layer can be formulated by setting $i_*=0$ while keeping $q>0$. The source term $q/R$ remains and radiation can escape through both boundaries. In that case $\mathcal H$ is undefined because $h_*=0$, so heating must be specified by $q$ rather than by $\mathcal H$. If both $i_*$ and $q$ vanish, the zero-radiation solution has $h_0=0$ and constant $\beta$, with $R$ determined by equation~(\ref{eq:mass}). The numerical survey below concerns illuminated layers; it is not a survey of the unilluminated family.

\section{Solving the coupled layer}
\subsection{Angular transport}
We discretize equation~(\ref{eq:conservative}) with conservative upwind finite volumes. Composite Gauss--Legendre quadrature on $[-1,0]$, $[0,\eta_b]$, $[\eta_b,\eta_s]$ and $[\eta_s,1]$ resolves the boundary characteristic sectors. At a trial velocity profile, equation~(\ref{eq:source}) produces a linear angular redistribution system, solved by sparse direct factorization. We then integrate equation~(\ref{eq:flow}) and damp the velocity update by one half. Iteration stops when the maximum velocity update is below $2\times10^{-9}$.

The reference models use 320 spatial cells and 96 angular nodes. Column-matched models use 256 cells and 80 nodes. Smooth surface-ray diagnostics are integrated independently through shape-preserving interpolants of the converged profiles, using DOP853 with relative tolerance $2\times10^{-10}$. Reported angular contrasts use $\mu\geq0.1>\eta_s$, while the coupled source retains the full angular range. Independent spatial and angular refinements, analytic fixed-coefficient tests and conservation budgets are included in the accompanying code.

\subsection{Frozen atmospheres}
For each coupled evolving solution (TD), we calculate two frozen radiation controls on exactly its $R(\eta)$ and $\beta(\eta)$ profiles. Both retain $D$ and use
\begin{equation}
\mu i'=R(D^3s_0-D^{-1}i).\label{eq:frozen}
\end{equation}
The fixed-source control (FS) retains the evolving $s_0(\eta)$ and isolates the formal transport change. The relaxed-source control (FR) recalculates $j_0$ and $s_0=j_0+q/R$ while retaining the material profiles and heating. FR measures the response of the radiation source to freezing; it is not a separate steady hydrodynamical solution.

The same illumination function is imposed on incoming directions. Frozen inflow sectors are $\mu>0$ at the bottom and $\mu<0$ at the top, whereas evolving inflow sectors depend on the boundary speeds. Thus the controls include the change in characteristic boundary classification. In FR this change can affect the source through angular redistribution, so FR does not isolate an interior time-derivative term independently of the boundary treatment. The reported surface rays, with $\mu\geq0.1$, have identical prescribed lower illumination in all controls.

We summarize angular shape by
\begin{equation}
C=I_s(0.1)/I_s(1),\quad
\delta C_{\rm FS}=100(C_{\rm TD}/C_{\rm FS}-1).\label{eq:contrast}
\end{equation}
Here $C>1$ means oblique emission exceeds normal emission at the specified angle; it does not characterize the unreported limit $\mu\rightarrow\eta_s$. A separate diagnostic sets $D=1$ on the same profiles and source. That diagnostic measures sensitivity to velocity coefficients, and is not a dynamically consistent non-relativistic solution.

\begin{figure*}
\centering\includegraphics[width=0.90\textwidth]{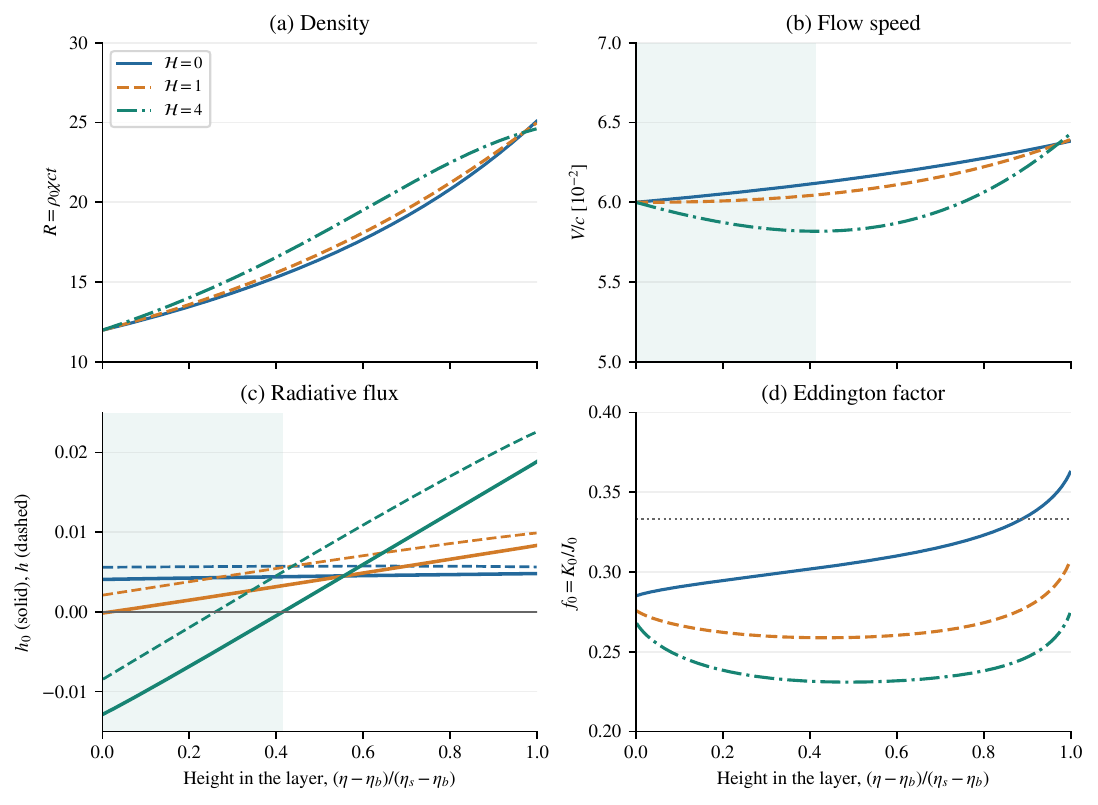}
\caption{Radiative heating and the flow. Density, velocity, radiative flux and Eddington factor across the layer are shown for $\mathcal H=0,1,4$ at $C_m=0.6$. In panel (c), solid curves show the comoving flux $h_0$ and dashed curves the laboratory flux $h$; colours identify the heating strength as in panel (a). The shaded region marks downward comoving flux in the strongly heated model, where the gas initially decelerates. The dotted $f_0=1/3$ line is a reference, not a closure. Alt text: Four line plots show density, velocity, radiative flux, and the Eddington factor across the layer for three heating strengths. Both fluxes are initially negative in the strongly heated flow; the laboratory flux changes sign below the comoving flux. The flow first decelerates and then accelerates.}\label{fig:structure}
\end{figure*}
\begin{figure*}
\centering\includegraphics[width=0.90\textwidth]{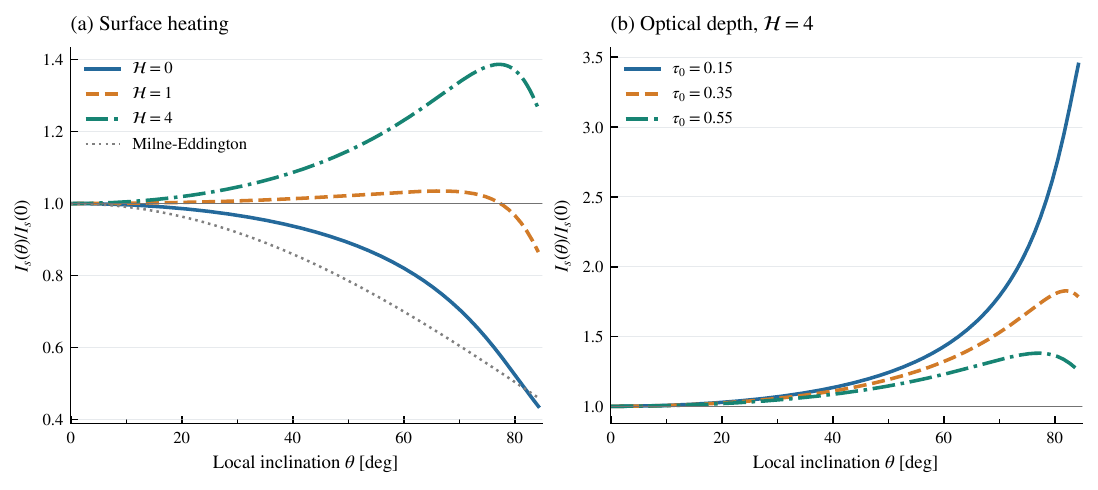}
\caption{From limb darkening to limb brightening. Emergent intensity is shown against local inclination $\theta=\arccos\mu$. Left: increasing heating at fixed mass flux. Right: varying column at $\mathcal H=4$. The static Milne--Eddington shape is shown for orientation; the grey horizontal line marks equal oblique and normal intensity. Alt text: Two panels show angular intensity curves for three heating strengths and three optical columns, with an approximate static limb darkening law for comparison. Strong heating increases the relative brightness at oblique angles.}\label{fig:angular}
\end{figure*}

\section{Emergent radiation}
\subsection{Heating and limb brightening}
Figure~\ref{fig:structure} shows the coupled flow and angular moments. The surface speed is about $0.064c$, compared with the injection speed $0.06c$. Internal heating produces radiation that can escape through either boundary. In the $\mathcal H=4$ model, downward emission from the layer outweighs the prescribed illumination near the base, so both the laboratory and comoving net fluxes are initially negative. Uniform positive heating constrains the change of the energy flux through equation~(\ref{eq:energy}); it does not prescribe a positive flux at every height. A flux can increase upwards while remaining negative below its zero crossing. The lower boundary is illuminated but is neither reflecting nor constrained to have zero outgoing radiation.

The radiation force is controlled by $h_0$, which is related to the laboratory moments by
\begin{equation}
h_0=\gamma^2[(1+\beta^2)h-\beta(j+k)].\label{eq:fluxframes}
\end{equation}
For $\mathcal H=4$, the computed zero crossings are approximately $\eta=0.0178$ for $h$ and $\eta=0.0225$ for $h_0$. Between these heights the laboratory flux is upward but the comoving flux is downward. The gas therefore continues to decelerate until $h_0$ changes sign, reaching a minimum speed of about $0.0582c$, and subsequently accelerates. Figure~\ref{fig:structure}(c) displays both fluxes. The negative comoving flux near the base is thus not solely a frame-transformation effect; the laboratory flux is negative there as well.

The outward rise of density also follows from the evolving geometry. Continuity conserves $\gamma R(\beta-\eta)$, not the stationary-flow quantity $\gamma R\beta$. Differentiating equation~(\ref{eq:mass}) gives
\begin{equation}
\frac{R'}{R}=\frac{1-\gamma^2(1-\beta\eta)\beta'}{\beta-\eta}.
\label{eq:densitygradient}
\end{equation}
On the selected branch, $R$ increases whenever $\beta'<(1-\beta^2)/(1-\beta\eta)$, including a range of positive velocity gradients. This inequality holds throughout all three reference profiles. Although $\beta$ rises in the accelerating region, the relative speed $\beta-\eta$ decreases sufficiently rapidly for $R$ to rise. This is a spatial density gradient at fixed time, not necessarily compression of a gas parcel. Indeed, along a parcel, $(\partial_t+\beta c\partial_z)\ln\rho_0=-\gamma^2(1-\beta\eta)\beta'/t$; the density of an accelerating parcel decreases even where the instantaneous profile rises outwards. The density and velocity profiles therefore satisfy the time-dependent continuity equation without the inverse relation familiar from a steady, constant-area flow.

At $\mathcal H=0$, the reference layer has $\tau_0=0.513$ and $C=0.436$. At $\mathcal H=1$, $C=0.865$; at $\mathcal H=4$, $C=1.264$ and $\tau_0=0.545$. Heating therefore changes the layer from obliquely dim to obliquely bright (Fig.~\ref{fig:angular}). Oblique paths intercept more of the distributed emission in a translucent layer, while an unheated illuminated layer preferentially transmits radiation along shorter paths. The frozen heated models also brighten towards the limb, consistent with static finite-depth calculations \citep{FukueAkizuki2006,Samadi2020}. Evolution changes this underlying angular pattern rather than creating limb brightening by itself.

The comoving Eddington factor near the upper boundary changes from approximately $0.363$ to $0.274$ across this sequence. These values are numerical angular moments at the outermost cell, rather than an imposed surface condition. Their variation demonstrates the departure from a constant Eddington factor in this family.

\subsection{A reversal with optical depth}
\begin{table}
\centering
\caption{Column-matched models at $\mathcal H=4$. The final column is a percentage; all other entries are dimensionless.}
\label{tab:columns}
\begin{tabular}{@{}rrrrr@{}}
\toprule
$\tau_0$ & $C_m$ & $C_{\rm TD}$ & $C_{\rm FS}$ & $\delta C_{\rm FS}$ \\
\midrule
0.15 & 0.1680 & 3.4482 & 2.9031 & +18.78 \\
0.35 & 0.3883 & 1.7843 & 1.7554 & +1.65 \\
0.55 & 0.6049 & 1.2561 & 1.2887 & -2.53 \\
\bottomrule
\end{tabular}
\end{table}

Table~\ref{tab:columns} compares the column-matched heated models. At $\tau_0=0.15$, evolution increases $C$ from $2.903$ to $3.448$, a correction of $18.8$ per cent. At $\tau_0=0.35$ the correction is $1.65$ per cent, while at $\tau_0=0.55$ it is $-2.53$ per cent. The relaxed-source controls give the same sign at these columns, with contrast corrections $19.2$, $1.80$ and $-2.72$ per cent, respectively. Thus the sign reversal is not an artifact of holding the source fixed.

Absolute intensity and normalized shape must be distinguished. In the thin model, $I_s(0.1)$ increases by $22.9$ per cent relative to FS, whereas $I_s(1)$ increases by $3.48$ per cent. In the reference fixed-mass heated model, both intensities rise, by $0.36$ and $2.92$ per cent respectively; their ratio consequently decreases. The atmospheric emissivity can thus increase at every reported direction while becoming less obliquely weighted. This distinction matters when an atmosphere calculation supplies both the luminosity and the angular law of a disc element.

\begin{figure*}
\centering\includegraphics[width=0.90\textwidth]{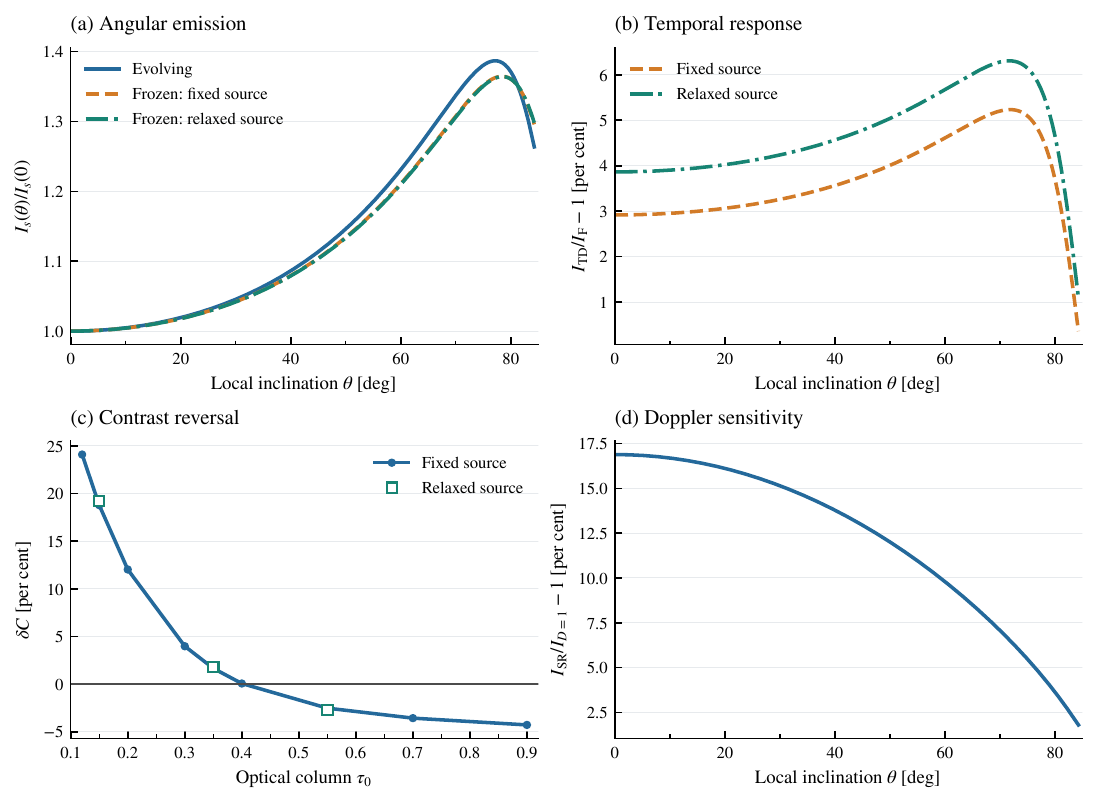}
\caption{How evolution changes the emerging light. Panels (a), (b) and (d) use the $C_m=0.6$, $\mathcal H=4$ layer: angular shape, temporal intensity corrections, and the formal sensitivity to velocity coefficients. Panel (c) follows the contrast correction through the matched-column sequence at $\eta_s=0.04$; open symbols show source relaxation. FS and FR retain the evolving material profiles. The $D=1$ comparison is a diagnostic at fixed source, not a second flow solution. Alt text: Four panels compare angular shapes, fractional intensity changes, the contrast correction across column, and sensitivity to velocity coefficients. Both frozen controls show a change in the sign of the contrast correction.}\label{fig:controls}
\end{figure*}

Figure~\ref{fig:controls} displays the angular changes and their dependence on column. The correction depends on how attenuation, distributed emission and the time-dependent intensity normalization act along each ray. The change of sign rules out a universal enhancement of oblique emission in this family. Whether a calibrated, angle-dependent effective optical depth can reproduce the result requires a separate test. The velocity-only diagnostic in the same figure also shows why a mildly relativistic flow must retain its Doppler coefficients when a temporal correction of a few per cent is being measured \citep{Krumholz2007}.

\begin{figure*}
\centering\includegraphics[width=0.90\textwidth]{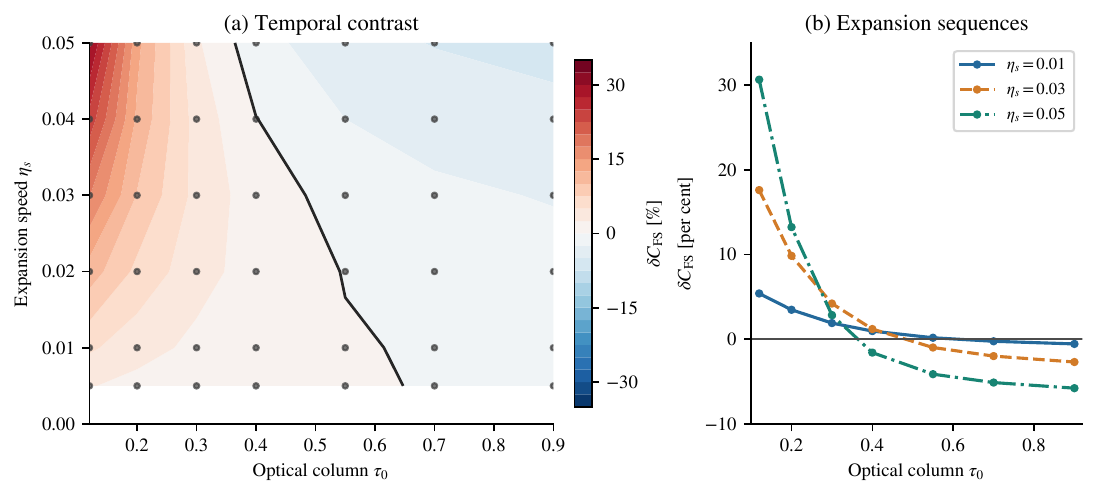}
\caption{The column-dependent reversal. Left: temporal contrast correction at $\mathcal H=4$ across column and expansion speed, with $\eta_b=\eta_s/4$. The colour scale is symmetric about zero; dots mark the 42 calculations and the heavy contour marks zero correction. Interpolation between samples is for display; the blank strip below $\eta_s=0.005$ lies outside the computed domain. Right: cuts through three computed expansion rates. Each column is matched by adjusting $C_m$. Alt text: A diverging colour map and three line curves show the temporal contrast correction across optical column and expansion speed. Positive corrections occupy the low column region and negative corrections appear at larger columns.}\label{fig:map}
\end{figure*}

\subsection{Why the contrast can change sign}
The fixed-source comparison admits an exact difference equation that clarifies the angular response. Write $\Delta i=i_{\rm TD}-i_{\rm FS}$ along a ray with $\mu>\eta_s$. Subtracting the two transfer equations at identical $R$, $D$, and $s_0$ gives
\begin{equation}
(\mu-\eta)\Delta i'+(R/D-1)\Delta i
=i_{\rm FS}+\eta i_{\rm FS}',\label{eq:difference}
\end{equation}
with $\Delta i(\eta_b)=0$. The right-hand side contains both the similarity intensity scaling and the spatial gradient sampled as the boundaries expand. Its cumulative effect depends on direction and column. Define the surface fractional intensity change $e_\mu=\Delta i_s(\mu)/i_{{\rm FS},s}(\mu)$. Then
\begin{equation}
\frac{C_{\rm TD}}{C_{\rm FS}}-1
=\frac{e_{0.1}-e_1}{1+e_1}.\label{eq:contrastdifference}
\end{equation}
For positive normal intensity, the contrast correction is negative whenever the fractional enhancement along the normal exceeds that along the oblique ray. Thus a reversal of angular contrast does not require a decrease in either absolute intensity. Equation~(\ref{eq:difference}) applies to FS; a relaxed source contributes an additional term $RD^3(s_{0,\rm TD}-s_{0,\rm FR})$ to the corresponding difference equation.

\subsection{Expansion and irradiation}
The 42-model survey in Fig.~\ref{fig:map} covers $0.12\leq\tau_0\leq0.9$ and $0.005\leq\eta_s\leq0.05$. Large positive corrections occur in thin, rapidly expanding layers; thicker models can have negative corrections. At the thin, fastest corner the correction reaches approximately $31$ per cent. At $\eta_s=0.05$, $\tau_0=0.9$, it is about $-5.9$ per cent after refinement. Expansion has its largest effect in the lower-column layers, where oblique photons sample a substantial source contribution without losing all memory of the incident field. The map quantifies this behavior for the specified heating and illumination.

\begin{figure*}
\centering\includegraphics[width=0.90\textwidth]{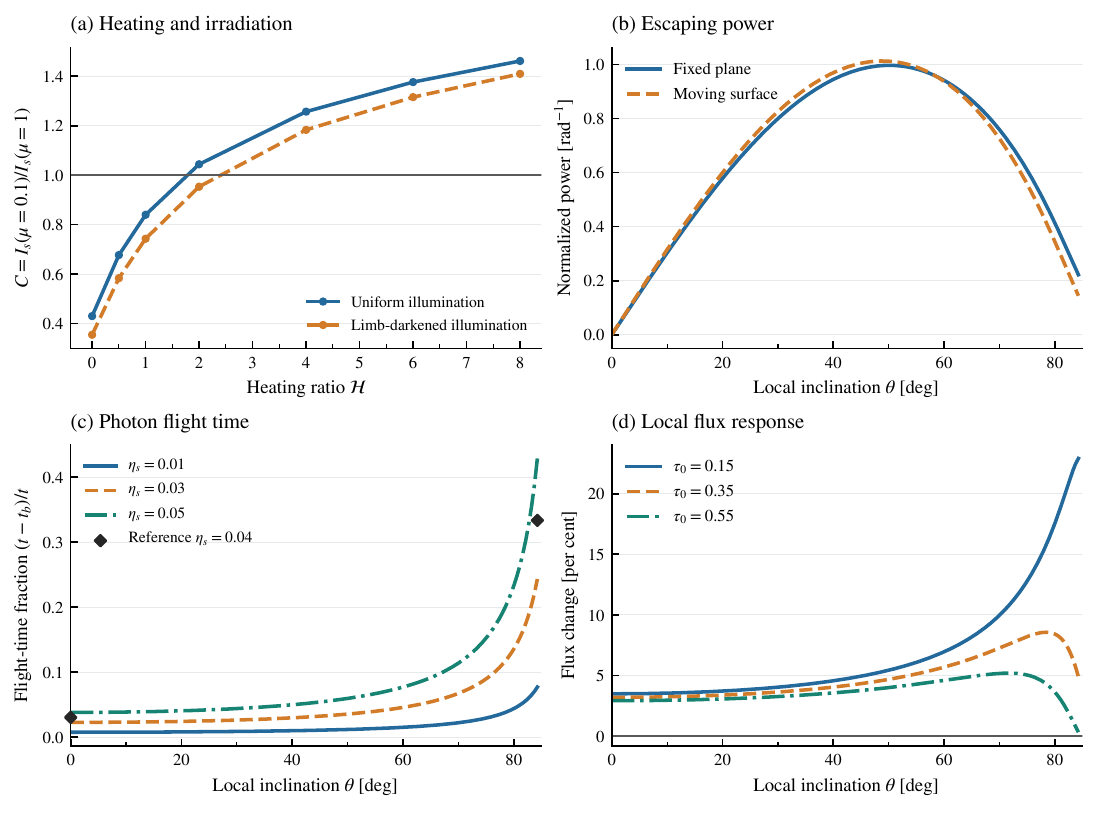}
\caption{Heating, viewing angle and photon travel. (a) Illumination dependence at matched $\tau_0=0.55$. (b) Power distributions for the reference heated layer, each normalized over $0.1\leq\mu\leq1$. (c) The flight-time fraction from equation~(\ref{eq:flight}) at three expansion rates; symbols mark the normal and $\mu=0.1$ reference rays. (d) The fractional change in the flux of an unobscured flat patch, equal to the TD--FS intensity correction at fixed projected area, for the three matched columns. This local diagnostic excludes global occultation and relativistic ray tracing. Alt text: Four plots compare illumination laws, angular power distributions, photon flight time fractions, and local flux corrections. Oblique rays sample longer time intervals than normal rays.}\label{fig:physical}
\end{figure*}

Figure~\ref{fig:physical} also compares $a=0$ and $a=2$ illumination at matched $\tau_0=0.55$. The latter is a diffusion-inspired angular law with the same reference positive-hemisphere illumination flux. Changing that law alters the heating strength at which oblique enhancement appears. The angular dependence of the emergent intensity therefore cannot be predicted from column and boundary speed alone; the incident field and the location of energy deposition remain necessary inputs.

\subsection{Accuracy of the emission}
Figure~\ref{fig:validation} collects independent checks. The analytic transfer tests in Appendix~\ref{app:analytic} show spatial convergence for both evolving and frozen transport. Refining the coupled reference solution from 320 to 640 cells changes $C$ by about $0.03$ per cent; the angular refinement is considerably smaller for the chosen composite quadrature. Selected map refinements from $96\times48$ to $256\times80$ change $\delta C_{\rm FS}$ by at most $0.081$ percentage points among the four tested locations. The map is adequate to establish the sign structure, but not to locate a zero contour to arbitrary precision.

In the reference sequence, the maximum integrated energy residual normalized by $h_*+q(\eta_s-\eta_b)$ is below $2\times10^{-8}$, and the momentum residual is below $3\times10^{-7}$ using the same normalization. Sparse-system residuals are of order $10^{-14}$ and the converged intensities are non-negative. These conservation residuals measure internal numerical consistency, not the physical error introduced by the grey, cold-flow assumptions. The quoted contrast digits in the table facilitate reproduction; physical interpretation does not require their last decimal.

\begin{figure*}
\centering\includegraphics[width=0.90\textwidth]{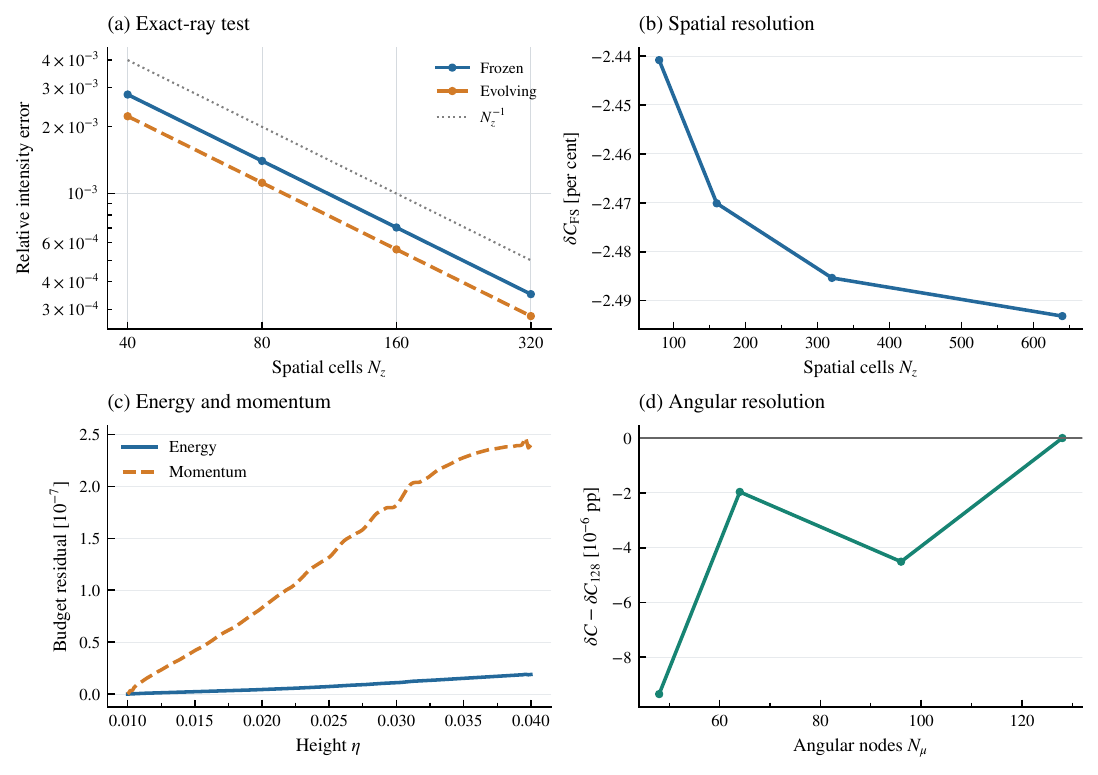}
\caption{Accuracy of the angular solution. (a) Finite-volume errors against exact rays, with a first-order reference. (b) Spatial convergence of the temporal contrast. (c) Integrated radiation--matter energy and momentum residuals, normalized by the supplied radiative flux. (d) Angular changes relative to the 128-node result at fixed spatial resolution. These are numerical checks within the specified physical model. Alt text: Four numerical validation plots show analytic ray errors, spatial convergence, conservation residuals, and angular refinement. Errors decrease with refinement while conservation residuals remain small.}\label{fig:validation}
\end{figure*}

\section{Implications for accretion flows}
\subsection{Photon flight times}
For a photon reaching the upper boundary at time $t$, the lower-boundary crossing occurred at $t_b=t(\mu-\eta_s)/(\mu-\eta_b)$, provided $\mu>\eta_s$. Its travel time is therefore
\begin{equation}
\frac{t-t_b}{t}=\frac{\eta_s-\eta_b}{\mu-\eta_b}.\label{eq:flight}
\end{equation}
In the reference layer this fraction is $0.333$ at $\mu=0.1$ and $0.0303$ at $\mu=1$. Figure~\ref{fig:physical} shows how rapidly the sampled time interval grows towards high local inclination. The photon thus samples an evolving atmosphere even when its vertical thickness is only $0.03ct$. Both density and intensity scale as $t^{-1}$ at fixed $\eta$, while the integrated column stays constant. Equation~(\ref{eq:flight}) is a kinematic measure of source evolution sampled along a ray, not a prediction of a Fourier lag or a diffusion residence time.

The dimensional scale remains free. For a layer of thickness $L=ct(\eta_s-\eta_b)$,
\begin{equation}
t=0.111\left(\frac{L}{10^8\,{\rm cm}}\right)
\left(\frac{\eta_s-\eta_b}{0.03}\right)^{-1}{\rm s}.
\end{equation}
This is a scaling relation, not an assignment of the layer to a particular source. It shows how the same angular calculation can be rescaled once a global model supplies a local thickness and evolution time. The fraction in equation~(\ref{eq:flight}) is a directional screening diagnostic. A frozen approximation additionally requires weak variation of the emission and extinction along the contributing characteristics; a short geometrical crossing time alone does not bound the error when spatial gradients or repeated scattering matter.

\subsection{From a patch to an observed flux}
For a flat, unobscured patch at distance $d$, with area $A_{\rm p}$ and laboratory viewing angle $\theta$, its bolometric flux contribution is $A_{\rm p}\mu I_s(\mu)/d^2$. At the same angle and projected area, the fractional TD--FS correction to this contribution equals the intensity correction. The thin-layer example therefore changes that local contribution by $22.9$ per cent at $\mu=0.1$, compared with $3.48$ per cent along the normal. The absolute projected contribution at very high inclination can nevertheless be small because it contains the factor $\mu$.

For an accretion disc, a local angular law must be propagated through the global geometry. In relativistic ray tracing, bolometric intensity transforms as $I_{\rm obs}=g^4I_{\rm em}$, where $g$ is the photon frequency-shift factor; integration is over the observer's image \citep{Li2005}. The laboratory frame used here already includes the vertical gas motion, so its emissivity must be transformed consistently into the frame used by that calculation. This avoids counting Doppler effects twice. The present solutions can supply a grey angular correction for such a calculation; predicting a spectral or black-hole-spin bias requires a frequency-dependent atmosphere \citep{Davis2006} and the complete image integration.

The flux relevant to radiation--matter conservation is distinct from the flux seen through a fixed plane. Across the upper moving boundary it is $h-\eta_sj$, whereas through a laboratory plane it is $h$. Likewise, the distributions per polar angle are proportional to $I_s(\cos\theta-\eta_s)\sin\theta$ and $I_s\cos\theta\sin\theta$, respectively. Figure~\ref{fig:physical} displays both measures. The boundary condition $\mu>\eta_s$ specifies outward crossing of this chosen surface; it does not impose a universal observer escape cone or identify the boundary speed with the material velocity.

\subsection{Disc atmospheres and reprocessing gas}
The nearest physical application is an already moving, locally plane-parallel surface layer whose heating and density change during photon transport. Dissipation near disc surfaces \citep{JiangCorona2014} provides motivation for examining heated, low-column gas. Our heating parameter measures the radiated local input relative to illumination from below. It is not an accretion-rate prescription: a disc model must determine how the dissipated energy is partitioned between the gas, radiation and magnetic field.

At super-Eddington accretion rates, outflow geometry and radiative transport jointly affect the emerging light \citep{Poutanen2007,JiangGlobal2014}. The solution isolates a local temporal contribution that could be tested on patches of an outflow or funnel wall. Its relevance depends on whether the curvature, transverse transport and omitted forces are small over that patch. It does not compute the global collimation or the mass-loss rate of the wind.

Tidal disruption events provide a related context. The models of \citet{Dai2018} and \citet{Thomsen2022} connect inclination and changing accretion conditions with the reprocessed emission. Multidimensional transfer also changes the available escape paths and ionization structure \citep{Parkinson2025}. Our result suggests a further question for such calculations: whether the local emissivity can be treated as frozen during photon passage through a changing surface layer. It does not determine an optical-to-X-ray ratio, since the present opacity is grey. Similarly, the angular structure is relevant motivation for polarized transfer, but predicting the polarization measured in accreting systems \citep{Krawczynski2022} requires a scattering phase matrix and Stokes transport.

\subsection{Relation to established solutions}
The approximate Milne--Eddington angular law, $I(\mu)/I(1)=(\mu+2/3)/(1+2/3)$, supplies the optically thick static reference in Fig.~\ref{fig:angular}. Finite-depth heated solutions \citep{FukueAkizuki2006,Samadi2020} are more appropriate comparisons for the underlying limb brightening. Their column, source and boundary data differ from a semi-infinite atmosphere, as do ours. We therefore use the classical law to identify a familiar angular shape, and use matched frozen transfer to measure the temporal correction quantitatively.

The numerical source and force are calculated from the full angular field, extending the present comparison beyond a prescribed Eddington factor. This follows the established motivation of variable angular closure \citep{Auer1970,Fukue2014}. The steady coupled flows of \citet{Fukue2015} and the temporal plane-parallel similarity solutions of \citet{Fukue2018} are the closest dynamical precedents. Fukue (2018) already established grey, time-dependent plane-parallel radiation-hydrodynamical similarity flows using comoving moment equations. The present advance is therefore not the introduction of that similarity geometry. Here we solve the angular transfer equation, evaluate its moments directly, and use matched frozen controls to quantify the emergent angular correction. The reported solutions do not establish continuation through, or removal of, the singular loci encountered in the earlier moment treatment. Relative to prescribed-flow radiation tests \citep{Lucy2005}, the velocity here is a coupled unknown. The contribution of the present family is a resolved temporal angular response, including its reversal with column, supported by matched controls and conservation budgets. These comparisons place the solution between steady atmosphere calculations and global time-dependent flows.

\subsection{Domain of the model}
Gravity, gas pressure and magnetic stresses are absent, so the gas enters with a prescribed velocity and the solution applies above an unspecified launching region. A physical application must check that the omitted forces do not appreciably alter the flow over the layer. Near a reversal of the radiative flux, radiative dominance cannot simply be assumed. The imposed column and upper boundary also do not locate a photosphere dynamically.

Prompt radiative balance neglects thermal storage, and the isotropic grey source excludes frequency redistribution and polarized Thomson scattering. These approximations make the coupling tractable but limit its spectral interpretation. The solution is a similarity branch for $t>0$, rather than an initial-value calculation or a demonstration of stability. Within this domain it provides both an angular emissivity calculation and a conservation test for more general time-dependent atmosphere and outflow solvers.

\subsection{Physical scope of the similarity evolution}\label{sec:similarity}
The prescribed time dependences are a restriction on the physical history, not merely a change of variables. At fixed similarity coordinate, $\rho_0\propto t^{-1}$ and $I\propto t^{-1}$, while the thickness increases as $t$. Consequently, the opacity column $\tau_0$ and the laboratory rest mass per unit area, $\chi^{-1}\int_{\eta_b}^{\eta_s}\gamma R\,d\eta$, remain constant. The layer is nevertheless an open flow: gas crosses both boundaries, with equal rest-mass fluxes per unit area $C_m/(\chi t)$ relative to those boundaries. Constant column does not mean that the same gas parcels remain inside the layer.

The heating prescription is equally specific. Uniform $q$ gives a proper volumetric heating rate $q^+\propto t^{-2}$ and, at fixed $\eta$, a specific heating rate $q^+/\rho_0=c^2q/(tR)\propto t^{-1}$. The corresponding laboratory heating power per unit area is $c^2(\chi t)^{-1}\int_{\eta_b}^{\eta_s}\gamma q\,d\eta$, which decreases as $t^{-1}$, as does the imposed illumination flux. The ratio $\mathcal H$ is therefore constant. The model describes neither heating held constant in physical units nor an arbitrary transient luminosity history.

A possible local application is an expanding surface or reprocessing layer during an interval in which its column changes slowly, its thickness grows approximately linearly, and its incident and dissipated powers decline together. These conditions must be checked against a global calculation or inferred evolution; they are not predicted here. In a heated disc atmosphere or radiative outflow, the practical insight is that nearly constant column alone does not justify stationary transfer: density and emissivity can change appreciably during the longer passage of oblique photons. The matched frozen calculations isolate that consequence within a specified evolution. Extrapolation to a steady disc, the launch of a wind, or a transient with a different heating history requires a more general calculation. The similarity solution applies over $t>0$ and does not describe the singular limit $t\rightarrow0$ or determine how the assumed evolution is established.

\section{Conclusions}
An evolving accretion layer can change its angular emissivity while retaining the same integrated optical column. Our coupled similarity solutions follow this response with the source and radiative acceleration determined by one angular radiation field. They show that heating sets the underlying limb-darkened or limb-brightened regime, while temporal transport changes both the intensity and its angular contrast.

At $\mathcal H=4$, the evolving oblique-to-normal contrast exceeds a matched frozen, fixed-source result by $18.8$ per cent at $\tau_0=0.15$ and falls below it by $2.53$ per cent at $\tau_0=0.55$. Source relaxation in the frozen layer preserves the sign change. The effect therefore cannot be described by a universal enhancement of limb brightening.

The physical timescale is the photon passage time at the relevant angle. In the reference layer, an oblique ray samples one third of the age of the flow even though the vertical crossing time is only three per cent of that age. Local emissivity prescriptions for rapidly changing disc surfaces or reprocessing gas can consequently require temporal transfer before a global luminosity or spectrum is calculated. The conservation derivation and reproducible angular solutions provide a basis for testing that step and for adding the thermal physics and geometry needed by specific accreting systems.

%\section*{Acknowledgements}

\section*{Data availability}
The numerical solver and supporting materials used to produce the model results and figures are available upon request, including the standalone notebook and reproduction-check files.

\appendix
\section{Radiation and matter conservation}\label{app:conservation}
We derive equations~(\ref{eq:mass})--(\ref{eq:momentum}) in the conventions used here. The covariant formulation follows \citet{Mihalas1984}; the reduction below specifies the cold-flow and heating assumptions needed for this particular similarity solution. The coordinate $x^0=ct$, the four-velocity $u^\alpha=\gamma(1,0,0,\beta)$ and the metric has signature $(-,+,+,+)$.

\subsection{Mass flux}
The continuity equation $\partial_\alpha(\rho_0u^\alpha)=0$ reads
\begin{equation}
\partial_t(\gamma\rho_0)+c\partial_z(\gamma\rho_0\beta)=0.
\end{equation}
Writing $Q=\gamma R$, the $t^{-1}$ density scaling gives $-Q-\eta Q'+(Q\beta)'=0$, or
\begin{equation}
[Q(\beta-\eta)]'=0.
\end{equation}
Its first integral is $C_m=\gamma R(\beta-\eta)$, equation~(\ref{eq:mass}). This is the flux relative to a surface of fixed $\eta$; the laboratory flux $\gamma R\beta$ is not constant along the layer.

\subsection{Radiative acceleration}
The material stress tensor is $T_{\rm m}^{\alpha\nu}=\rho_0c^2u^\alpha u^\nu$. In the comoving frame, heating and prompt isotropic re-emission cancel in the material energy equation. The remaining material four-force density has components
\begin{equation}
G_{{\rm m},0}^{\alpha}=(0,0,0,\rho_0\chi F_0/c).
\end{equation}
Thus $u_\alpha G_{\rm m}^{\alpha}=0$, as required for a pressureless gas of fixed rest mass. A boost gives $G_{\rm m}^{z}=\gamma\rho_0\chi F_0/c$. Mass conservation reduces $\partial_\nu T_{\rm m}^{\alpha\nu}=G_{\rm m}^{\alpha}$ to $\rho_0c^2u^\nu\partial_\nu u^\alpha=G_{\rm m}^{\alpha}$. Its vertical component is
\begin{equation}
(\partial_t+\beta c\partial_z)(\gamma\beta)=\frac{\chi F_0}{c^2}.
\end{equation}
Since $F_0=c^2h_0/(\chi t)$ and $d(\gamma\beta)/d\beta=\gamma^3$, the left-hand side becomes $\gamma^3(\beta-\eta)\beta'/t$. Hence
\begin{equation}
\gamma^3(\beta-\eta)\beta'=h_0,
\end{equation}
which is equation~(\ref{eq:flow}). The power of $\gamma$ follows from differentiating the momentum per unit rest mass, rather than the velocity alone.

\subsection{Radiation exchange}
Let $\langle X\rangle=\frac12\int_{-1}^{1}X\,d\mu$. Frequency-integrated intensity and solid angle transform as $i=D^4i_0$ and $d\mu=D^{-2}d\mu_0$. Using $D=\gamma(1+\beta\mu_0)$ gives
\begin{align}
\langle D^3\rangle&=\gamma,&\langle\mu D^3\rangle&=\gamma\beta,\\
\langle D^{-1}i\rangle&=\gamma(j_0+\beta h_0),\\
\langle\mu D^{-1}i\rangle&=\gamma(\beta j_0+h_0).
\end{align}
The zeroth and first angular moments of the conservative transfer equation are consequently
\begin{align}
(h-\eta j)'&=\gamma R(s_0-j_0-\beta h_0),\\
(k-\eta h)'&=\gamma R[\beta(s_0-j_0)-h_0].
\end{align}
Substituting $R(s_0-j_0)=q$ yields equations~(\ref{eq:radenergy}) and (\ref{eq:radmomentum}). Positive $h_0$ transfers upward momentum to the gas, so its sign in the radiation momentum equation is negative. The heating terms $\gamma q$ and $\gamma\beta q$ are the laboratory energy and momentum supplied by an isotropic comoving energy source.

\subsection{Combined budgets}
From $\gamma'=\gamma^3\beta\beta'$ and equation~(\ref{eq:flow}),
\begin{align}
[C_m(\gamma-1)]'&=\gamma R\beta h_0,\\
[C_m\gamma\beta]'&=\gamma Rh_0.
\end{align}
Adding these expressions to the radiation moments cancels the internal exchange and gives equations~(\ref{eq:energy}) and (\ref{eq:momentum}). The integrated tests in Fig.~\ref{fig:validation} subtract the base value and the integral of the external heating term from each total flux. In the unheated limit both totals are constant. These identities assume that heating leaves no stored material internal energy; including a thermal equation would add enthalpy and pressure terms to the material fluxes.

\section{Exact ray solutions}\label{app:analytic}
For a verification problem with constant $R$, $\beta$ and $s_0=s$, $D$ is constant along each ray. Set $k=R/D-1$ and $A=RD^3s/k$. For $\mu>\eta_s$ and $k\ne0$, equation~(\ref{eq:transfer}) integrates to
\begin{equation}
i_s=A+(i_b-A)\left(\frac{\mu-\eta_s}{\mu-\eta_b}\right)^k .
\end{equation}
At $k=0$, its continuous limit is $i_s=i_b+RD^3s\ln[(\mu-\eta_b)/(\mu-\eta_s)]$. The frozen solution is
\begin{equation}
i_s=D^4s+(i_b-D^4s)\exp\left[-\frac{R(\eta_s-\eta_b)}{D\mu}\right].
\end{equation}
These tests prescribe the source and do not claim to solve the coupled material equations. Figure~\ref{fig:validation} uses $R=12$, $\beta=0.08$, $s=0.02$ and the reference domain, with 40--320 spatial cells. The independently integrated emergent rays used for the science figures do not replace the finite-volume angular field used to close the source.
\bibliographystyle{mnras}
\bibliography{references}
\end{document}